\documentclass[12pt,aps,prd,preprint,showpacs]{revtex4-1}

\usepackage[utf8]{inputenc}
\usepackage[T1]{fontenc}
\usepackage[pdftex]{hyperref}
\usepackage{amsmath,amsfonts,amssymb,amsthm,graphicx,graphics,float}

\begin{document}

\title{Determination of the $\theta_{23}$ octant in long baseline neutrino experiments within and beyond the standard model}

\author{C. R. \surname{Das}}
\email[]{das@theor.jinr.ru}
\affiliation{Bogoliubov Laboratory of Theoretical Physics, Joint Institute of Nuclear Research, Joliot-Curie 6, 141980 Dubna, Moscow region, Russia}

\author{Jukka \surname{Maalampi}}
\email[]{jukka.maalampi@jyu.fi}
\affiliation{University of Jyvaskyla, Department of Physics, P.O.\,Box 35, FI-40014 University of Jyvaskyla, Finland}

\author{Jo\~ao \surname{Pulido}}
\email[]{pulido@cftp.ist.utl.pt}
\affiliation{Centro de F\'isica Te\'orica das Part\'iculas, Instituto Superior T\'ecnico (CFTP-IST), Av.\,Rovisco Pais, P-1049-001 Lisboa, Portugal}

\author{Sampsa \surname{Vihonen}}
\email[]{sampsa.p.vihonen@student.jyu.fi}
\affiliation{University of Jyvaskyla, Department of Physics, P.O.\,Box 35, FI-40014 University of Jyvaskyla, Finland}

\date{\today}

\begin{abstract}
The recent data indicate that the neutrino mixing angle $\theta_{23}$ deviates from the maximal-mixing value of 45$^\circ$, showing two nearly degenerate solutions, one in the lower octant (LO) ($\theta_{23}<45^\circ$) and one in the higher octant (HO) ($\theta_{23}>45^\circ$). We investigate, using numerical simulations, the prospects for determining the octant of $\theta_{23}$ in the future long baseline oscillation experiments. We present our results as contour plots on the ($\theta_{23}-45^\circ$, $\delta$)--plane, where $\delta$ is the $CP$ phase, showing the true values of $\theta_{23}$ for which the octant can be experimentally determined at 3$\,\sigma$, 2$\,\sigma$ and 1$\,\sigma$ confidence level. In particular, we study the impact of the possible nonunitarity of neutrino mixing on the experimental determination of $\theta_{23}$ in those experiments.
\end{abstract}

% PACS numbers (Neutrino oscillations, nonstandard oscillations)

\maketitle

\section{\label{Introduction}Introduction}

Many solar, atmospheric, reactor and accelerator neutrino experiments have firmly established the existence of neutrino oscillations. Neutrino oscillations can be parametrized in terms of six physical variables, namely by three mixing angles $\theta_{12}$, $\theta_{23}$ and $\theta_{13}$, a phase $\delta_{CP}$, and two squared-mass differences $\Delta m_{21}^2=m_2^2-m_1^2$ and $\Delta m_{31}^2=m_3^2-m_1^2$. These parameters are by now experimentally quite precisely determined, with the exception of the $CP$ phase $\delta_{CP}$. As to the mixing angle $\theta_{23}$, one still do not know in which octant it lies ($\theta_{23}<45^\circ$ or $\theta_{23}>45^\circ$). Also the order of the masses of three light neutrinos ($\nu_1$, $\nu_2$, $\nu_3$) remains unknown, namely whether it is $m_3\geq m_1,m_2$ (normal hierarchy, NH) or $m_3\leq m_1,m_2$ (inverted hierarchy, IH).

As to the mixing angle $\theta_{23}$, also known as the atmospheric angle, the fits on global data indicate that $\theta_{23}$ deviates from the maximal-mixing value 45$^\circ$ showing two degenerate solutions, a low-octant (LO) solution with $\theta_{23}^2<45^\circ$ and a high-octant (HO) solution with $\theta_{23}>45^\circ$ \cite{Fogli:1996pv,Forero:2014bxa,Capozzi:2016rtj,Gonzalez-Garcia:2015qrr}. This octant degeneracy is one of many parameter degeneracies that hamper the interpretation of neutrino oscillation data \cite{Fogli:2012ua}. The NO$\nu$A experiment has recently excluded the maximal-mixing value $\theta_{23}=45^\circ$ at the 2.6$\,\sigma$ confidence level \cite{Adamson:2017gxd}. Two statistically degenerate values for $\sin^2\theta_{23}$ were found, $0.404^{+0.030}_{-0.022}$ and $0.624^{+0.022}_{-0.030}$, which both explain the data on muon neutrino disappearance at the 68\% confidence level. Earlier experimental results are compatible with $\theta_{23}=45^\circ$ \cite{Wendell:2010md,Adamson:2014vgd,Abe:2015awa,Adamson:2016xxw}. The prospects of resolving the $\theta_{23}$ octant in next generation experiments have been studied in e.g.\,\cite{Agarwalla:2013ju,Ghosh:2014rna,Nath:2015kjg,Fukasawa:2016yue,Ballett:2016daj,Chatterjee:2017irl,Agarwalla:2017nld}.

Identifying the true value of $\theta_{23}$ is an important goal for future experiments, given its importance for understanding the mechanism behind neutrino masses and mixing. For example, one symmetry of the neutrino sector under the interchange of $\nu_\mu$ and $\nu_\tau$ would predict $\theta_{23}$ to have the maximal mixing value of 45$^\circ$ (see e.g.\,\cite{Altarelli:2010gt}). In that case, the $\nu_\mu$ and $\nu_\tau$ flavors would have an equal weight in the $\nu_3$ mass state. In some models the octant of $\theta_{23}$ is directly related to the neutrino mass hierarchy, e.g.\,in the model considered in \cite{CarcamoHernandez:2017owh}, the mass hierarchy for the higher octant is normal and for the lower octant the mass hierarchy is inverted.

In interpreting the data, one should take into account the possibility of beyond the standard model effects, which may appear as nonunitarity of the mixing of light neutrinos. In particular, the possible existence of sterile neutrinos and nonstandard neutrino interactions would influence the experimental determination of $\theta_{23}$ and the octant where the angle lies. The effects of some new physics in octant determination have been studied in future long baseline neutrino experiments in e.g.\,\cite{Hanlon:2013ska,Agarwalla:2016xlg,Agarwalla:2016fkh,Dutta:2016eks,Choubey:2017cba}.

In this paper, we will consider the determination of the mixing angle $\theta_{23}$ in long baseline neutrino experiments taking possible nonunitarity effects into account. First, we will update our previous studies \cite{Das:2016bwe, Das:2014fja}, done for the standard model case with three conventional neutrinos. We apply the results of the most recent fits for the values of mixing parameters and use as our benchmark the setup of the proposed Deep Underground Neutrino Experiment (DUNE). Our main goal is to find out how the possible nonunitarity of the mixing matrix of the light neutrinos affect the sensitivity of the experiments in identifying the octant of $\theta_{23}$. Such nonunitarity may arise, e.g., if one adds to the particle content of the standard model sterile neutrinos that mix with the conventional neutrinos. We also study to which extent the still unknown value of the $CP$ phase interferes with the determination of the octant of $\theta_{23}$, and we present our results for both the normal and inverted mass hierarchy.

This work is organized as follows. In Sec.\,\ref{Brief review of the formalism} we will present a brief review of the formalism we will use. In Sec.\,\ref{Numerical methods} the simulation method is described. The results of simulations are presented and discussed in Sec.\,\ref{Results}. The Sec.\,\ref{Conclusions} contains a summary and conclusions.

\section{\label{Brief review of the formalism}Brief review of the formalism}

We start our analysis by investigating the prospects for $\theta_{23}$ octant determination in long baseline experiments in the case of standard neutrinos. For these, the matter evolution is determined by the Hamiltonian which in the mass basis reads
\begin{widetext}
\begin{equation}
H=\frac{1}{2E}
\left(\begin{array}{ccc}
0 & 0 & 0\\
0 & \Delta m^2_{21} & 0\\
0 & 0 & \Delta m^2_{31}\end{array}\right)+U^{\dagger}\left(\begin{array}{ccc}V_{CC}+V_{NC} & 0 & 0\\
0 & V_{NC} & 0\\
0 &0 & V_{NC}
\end{array}\right)U.
\label{H}
\end{equation}
\end{widetext}
Here $U$ is the conventional neutrino mixing matrix (PMNS matrix), $V_{CC}=\sqrt{2}G_FN_e$ and $V_{NC}=-G_FN_n/\sqrt{2}$ are respectively the charged current and the neutral current matter potentials.

Owing to their smallness, the most common understanding of the origin of neutrino masses lies in the assumption of a new physics scale associated with the general seesaw mechanism, whereby heavy right-handed neutrinos are added to the particle content of the standard model. These $SU(3)_C\otimes SU(2)_L\otimes U(1)_Y$ singlet neutrinos mix with the standard neutrino flavors $\nu_e$, $\nu_\mu$, $\nu_\tau$ and, although in the likely case they are too heavy to be kinematically produced, they should leave traces at the energies within experimental reach. These traces appear in the oscillation probabilities through the $n\times n$ unitary matrix $\cal U$ connecting the neutrino mass and flavor eigenstates which generalizes the conventional $3\times 3$ $U$ matrix of the standard case. The $\cal U$ matrix can be written in the form \cite{Hettmansperger:2011bt}
\begin{equation}
\cal U=\left(\begin{array}{cc}
N & S\\
T & V
\end{array}\right)
\end{equation}
where $N$ and $S$ are ($3\times 3$) submatrices that contain respectively the mixing in the light (active) neutrino sector and the active-sterile mixing. Submatrices $T$ and $V$ define the mixing of the sterile states with the active and sterile states respectively. In this way, Eq.\,(\ref{H}) is modified to
\begin{widetext}
\begin{equation}
H=\frac{1}{2E}
\left(\begin{array}{ccc}
0 & 0 & 0\\
0 & \Delta m^2_{21} & 0\\
0 & 0 & \Delta m^2_{31}\end{array}\right)+N^{\dagger}\left(\begin{array}{ccc}V_{CC}+V_{NC} & 0 & 0\\
0 & V_{NC} & 0\\
0 & 0 & V_{NC}
\end{array}\right)N.
\label{HN}
\end{equation}
\end{widetext}

The unitarity of $\cal U$ implies that the matrix describing the mixing in the light sector, namely $N$, is no longer unitary. In \cite{Escrihuela:2015wra}, the matrix $N$ was presented in the form $N=N^{\rm NP}U$, where $U$ is the conventional unitary $3\times 3$ matrix and $N^{\rm NP}$ the triangular matrix
\begin{equation}
N^{\rm NP} =\left(\begin{array}{ccc}
\alpha_{11} & 0 & 0\\
\alpha_{21} & \alpha_{22} & 0\\
\alpha_{31} & \alpha_{32} & \alpha_{33}
\end{array}\right),
\label{NP1}
\end{equation}
parametrizing the deviations from unitarity. The description of the unitarity violation therefore requires three real parameters $\alpha_{ii}$ which are close to unity and three complex ones $\alpha_{ij}$ ($i\neq j$), which are close to zero.

A slightly different notation for the prefactor matrix $N^{\rm NP}$ was given in \cite{Blennow:2016jkn},
\begin{equation}
N^{\rm NP} =\left(\begin{array}{ccc}
1-\alpha_{ee} & 0 & 0\\
\alpha_{\mu e} & 1-\alpha_{\mu\mu} & 0\\
\alpha_{\tau e} & \alpha_{\tau\mu} & 1-\alpha_{\tau\tau}
\end{array}\right).
\label{NP2}
\end{equation}
Here $\alpha_{\ell\ell'}$ directly parametrize deviations from the unitarity and since these deviations are known to be small one has $\alpha_{\ell\ell}\ll 1$ and $\vert\alpha_{\ell\ell'}\vert\ll 1$. The difference between the parametrizations (\ref{NP1}) and (\ref{NP2}) is purely aesthetic, but as both of them are used parallel in the literature, we give them both here for convenience.

Since current experiments involve mainly electron and muon neutrinos, only $\alpha_{11}$, $\alpha_{22}$ and $\alpha_{21}$ (or equivalently $\alpha_{ee}$, $\alpha_{\mu\mu}$ and $\alpha_{\mu e}$) need to be considered, hence four new parameters are effectively required in the analysis. Constraints for $\alpha_{ij}$ and $\alpha_{\ell\ell'}$ are given in Refs.\,\cite{Blennow:2016jkn} and \cite{Escrihuela:2015wra}. No constraint exists, however, for the off-diagonal phases.

One should note that the nonunitary of the mixing of the neutrino flavors $\nu_e$, $\nu_\mu$ and $\nu_\tau$ would, in general, affect the determination of the mixing angles $\theta_{12}$, $\theta_{23}$ and $\theta_{31}$ from the existing neutrino oscillation data. However, in the triangular parametrizations of Eqs.\,(\ref{NP1}) and (\ref{NP2}) the nonunitarity effects disappear in the leading order and are hence negligible in comparison with the uncertainties of the experimental results used in the fits done \cite{Blennow:2016jkn}. Hence, the matrix $U$ has in good approximation the same numerical form as is obtained in the standard analysis of the data where the unitarity assumed to hold.

Our analysis, which follows the lines presented in \cite{Das:2016bwe}, is based on numerical simulations where we utilize the GLoBES software \cite{Huber:2004ka, Huber:2007ji}. Let us note that whereas analytical expressions for survival and conversion probabilities both in vacuum and matter have been given in the literature \cite{Akhmedov:2004ny, Nunokawa:2007qh}, for the nonunitarity effect in neutrino oscillations only the vacuum expressions exist \cite{Escrihuela:2015wra}. Brief discussions of the matter potential in the nonunitary case are given in Refs.\,\cite{Ge:2016xya} and \cite{Escrihuela:2016ube}.

\section{\label{Numerical methods}Numerical methods}

We evaluate the effect of nonunitary mixing on the determination of $\theta_{23}$ octant by simulating a long baseline oscillation experiment with the DUNE specifications with the GLoBES program. Since the nonunitary mixing matrix is not available in the standard GLoBES package, we modify the program by introducing our own add-on, which replaces the standard PMNS mixing matrix with its nonunitary version given by Eq.\,(\ref{NP1}), or alternatively Eq.\,(\ref{NP2}), and replaces the standard Hamiltonian shown in Eq.\,(\ref{H}) with its nonunitary version (\ref{HN}).

The octant discovery potential is evaluated for a given $\theta_{23}$ value as
\begin{equation}\label{ChiSquare}
\Delta\chi^2(\theta_{23})=\chi^2(90^\circ-\theta_{23})-\chi^2(\theta_{23}),
\end{equation}
where $\chi^2(\theta_{23})$ evaluates the chi-square distribution at the given true value $\theta_{23}$, whilst in $\chi^2(90^\circ-\theta_{23})$ it is evaluated at the wrong octant solution $90^\circ-\theta_{23}$. This leaves the subtraction of the two, $\Delta\chi^2$, an approximate chi-square distribution with one degree of freedom, and hence the sensitivity for ruling out the wrong octant at 1$\,\sigma$, 2$\,\sigma$ and 3$\,\sigma$ confidence levels is reached at $\Delta\chi^2=$ 1, 4, and 9, respectively.

The simulation of the DUNE setup is performed using the same experimental configuration that was used in the DUNE conceptual design report \cite{Acciarri:2015uup} and was published in Ref.\,\cite{Alion:2016uaj}. The octant discovery potential is evaluated using Eq.\,(\ref{ChiSquare}) as described above.

In this work, we assess the sensitivity to the $\theta_{23}$ octant in DUNE in four different scenarios. On the one hand, we update the octant sensitivity plots for the standard model case, where no sterile neutrinos exist and the oscillations would follow the standard three-neutrino paradigm. On the other hand, we also evaluate the octant sensitivity in scenarios, where sterile neutrinos do exist, manifesting themselves as nonunitarity of the mixing matrix of the three active neutrinos. The effects of sterile neutrinos to the oscillations would depend on the scale of the lightest sterile mass.

For the standard oscillation parameters, we employ the best-fit values and their associated errors, which have been obtained from the experimental data collected from the past and ongoing neutrino experiments (see Ref.\,\cite{Esteban:2016qun}). We take the central values and standard deviations of these parameter best-fits and take them into account as Gaussian distributions. For the mixing angles the Gaussian distributions are set for $\sin^2\theta_{12}$, $\sin^2\theta_{13}$ and $\sin^2\theta_{23}$. These values are presented in Table \ref{bounds:1}.\,\cite{notefoot}

In addition to the standard oscillation parameters, we also need to consider the new physics parameters $\alpha_{ij}$, $i,j=1,2,3$, and $\alpha_{\ell\ell'}$, $\ell,\ell'=e,\mu,\tau$, as indicated in Eqs.\,(\ref{NP1}) and (\ref{NP2}), respectively. Since no significant signs of physics beyond the standard model has been observed in oscillation experiments, there exist only upper bounds on the different $\alpha$ parameters. In this work we consider the possibility of sterile neutrino induced new physics by allowing the $\alpha_{ij}$ and $\alpha_{\ell\ell'}$ parameters to have Gaussian distributions, where central values are set at zero and standard deviations to match the appropriate upper bounds. The standard three-neutrino paradigm is restored by setting $\alpha_{11}$, $\alpha_{22}$, $\alpha_{33}=1$ and $\alpha_{21}$, $\alpha_{31}$, $\alpha_{32}=0$, or alternatively $\alpha_{\ell\ell'}=0$ for all $\ell,\ell'=e,\mu,\tau$ combinations.

\begin{table*}
\caption{\label{bounds:1}The experimental best-fit values and standard deviations for the standard neutrino oscillation parameters. These values are shown for both mass hierarchies and are taken from a recent global analysis \cite{Esteban:2016qun}. Note that $\Delta m_{3l}^2$ stands for $\Delta m_{31}^2$ in normal hierarchy (NH) and $\Delta m_{32}^2$ in inverted hierarchy (IH).}
\begin{ruledtabular}
\begin{tabular}{ccccc}
Parameter & Central value (NH) & Error (NH) & Central value (IH) & Error (IH)\\ \hline
$\sin^2\theta_{12}$ & 0.306 & 0.012 & 0.306 & 0.012\\
$\sin^2\theta_{13}$ & 0.02166 & 0.00075 & 0.02179 & 0.00076\\
$\sin^2\theta_{23}$ & 0.441 & 0.027 & 0.587 & 0.024\\
$\delta_{CP}$ ($^\circ$) & 261 & 59 & 277 & 46\\
$\Delta m_{21}^2$ (10$^{-5}$ eV$^2$) & 7.50 & 0.19 & 7.50 & 0.19\\
$\Delta m_{3l}^2$ (10$^{-3}$ eV$^2$) & 2.524 & 0.040 & -2.514 & 0.041
\end{tabular}
\end{ruledtabular}
\end{table*}

We start the investigation on the new physics scenarios by assuming that all three sterile neutrinos are too massive to be produced in the experiment. In this scenario, the sterile neutrinos do not contribute to the oscillations kinematically, but they affect the neutrino oscillation probabilities through the nonunitarity of the 3$\times$3 mixing matrix that controls the mixing of active neutrinos. These bounds have been evaluated for nonunitary mixing in two independent references; for the $\alpha_{ij}$ basis they are provided in Ref.\,\cite{Escrihuela:2016ube} and for $\alpha_{\ell\ell'}$ in Ref.\,\cite{Blennow:2016jkn}, and they are both shown in Table \ref{bounds:2}.

We also consider the scenario where at least one of the sterile neutrinos is sufficiently light to be produced kinematically in the experiment. In such case, sterile neutrinos could contribute to the oscillations in two different ways depending on their mass range. If the lightest sterile neutrino has its mass $m_4$ in the range such that $\Delta m_{41}^2\equiv m_4^2-m_1^2\sim 0.1-1$ eV$^2$, the active neutrinos could oscillate to the sterile neutrino state between the near and far detectors. It is also possible, however, that the oscillations to the sterile neutrino are too rapid to be observed in the far detector, i.e.\,they average out, but sufficiently light to occur before the near detector. In either case, not all constraints used in deriving the upper bounds in Table \ref{bounds:2} are applicable, and therefore new bounds must be derived. This topic has been thoroughly reviewed in Ref.\,\cite{Blennow:2016jkn}, where appropriate bounds were provided for $\alpha_{\ell\ell'}$, $\ell,\ell'=e,\mu,\tau$ in mass ranges $\Delta m_{41}^2\sim 0.1-1$ eV$^2$ and $\Delta m_{41}^2\geq 100$ eV$^2$. These bounds are shown in Table \ref{bounds:3}.

\begin{table*}
\caption{\label{bounds:2}Bounds on nonunitary parameters in both $\alpha_{ij}$ and $\alpha_{\ell\ell'}$ representations, taken from \cite{Escrihuela:2016ube} and \cite{Blennow:2016jkn}, respectively. The bounds are given in 90\% and 2$\,\sigma$ confidence levels.}
\begin{ruledtabular}
\begin{tabular}{cccc}
Parameter & Upper bound (90\% CL) & Parameter & Upper bound (2$\,\sigma$ CL)\\ \hline
$\alpha_{11}$ & $0.9974$ & $\alpha_{e e}$ & $1.3\times 10^{-3}$\\
$\alpha_{22}$ & $0.9994$ & $\alpha_{\mu\mu}$ & $2.2\times 10^{-4}$\\
$\alpha_{33}$ & $0.9988$ & $\alpha_{\tau\tau}$ & $2.8\times 10^{-3}$\\
$|\alpha_{21}|$ & $2.6\times 10^{-2}$ & $|\alpha_{\mu e}|$ & $6.8\times 10^{-4}$\\
$|\alpha_{31}|$ & $2.0\times 10^{-3}$ & $|\alpha_{\tau e}|$ & $2.7\times 10^{-3}$\\
$|\alpha_{32}|$ & $1.5\times 10^{-2}$ & $|\alpha_{\tau\mu}|$ & $1.2\times 10^{-3}$
\end{tabular}
\end{ruledtabular}
\end{table*}

\begin{table}
\caption{\label{bounds:3}Bounds on nonunitary parameters in $\alpha_{\ell\ell'}$ representation, taken from \cite{Blennow:2016jkn}. In this scenario the constraints would correspond to mixing with a light sterile neutrino in two mass scales: $\Delta m_{41}^2\sim 0.1-1$ eV$^2$ (left column) and $\Delta m_{41}^2\geq 100$ eV$^2$ (right column). The constraints are presented at a 95\% confidence level.}
\begin{ruledtabular}
\begin{tabular}{ccc}
Parameter & $\Delta m_{41}^2\sim 0.1-1$ eV$^2$ & $\Delta m_{41}^2\geq 100$ eV$^2$\\ \hline
$\alpha_{e e}$ & $1.0\times 10^{-2}$ & $2.4\times 10^{-2}$\\
$\alpha_{\mu\mu}$ & $1.4\times 10^{-2}$ & $2.2\times 10^{-2}$\\
$\alpha_{\tau\tau}$ & $1.0\times 10^{-1}$ & $1.0\times 10^{-1}$\\
$|\alpha_{\mu e}|$ & $1.7\times 10^{-2}$ & $2.5\times 10^{-2}$\\
$|\alpha_{\tau e}|$ & $4.5\times 10^{-2}$ & $6.9\times 10^{-2}$\\
$|\alpha_{\tau\mu}|$ & $5.3\times 10^{-2}$ & $1.2\times 10^{-2}$
\end{tabular}
\end{ruledtabular}
\end{table}

\section{\label{Results}Results}

We restrict our study to four different scenarios that could take place in the presence (or absence) of physics beyond the standard model. These scenarios include the standard three-neutrino paradigm, nonunitary neutrino mixing, mixing with a light sterile neutrino, and finally, a scenario where no constraints are set for the new physics parameters.

\subsection{\label{The standard model case}The standard model case}

\begin{figure*}
\includegraphics[width=\linewidth]{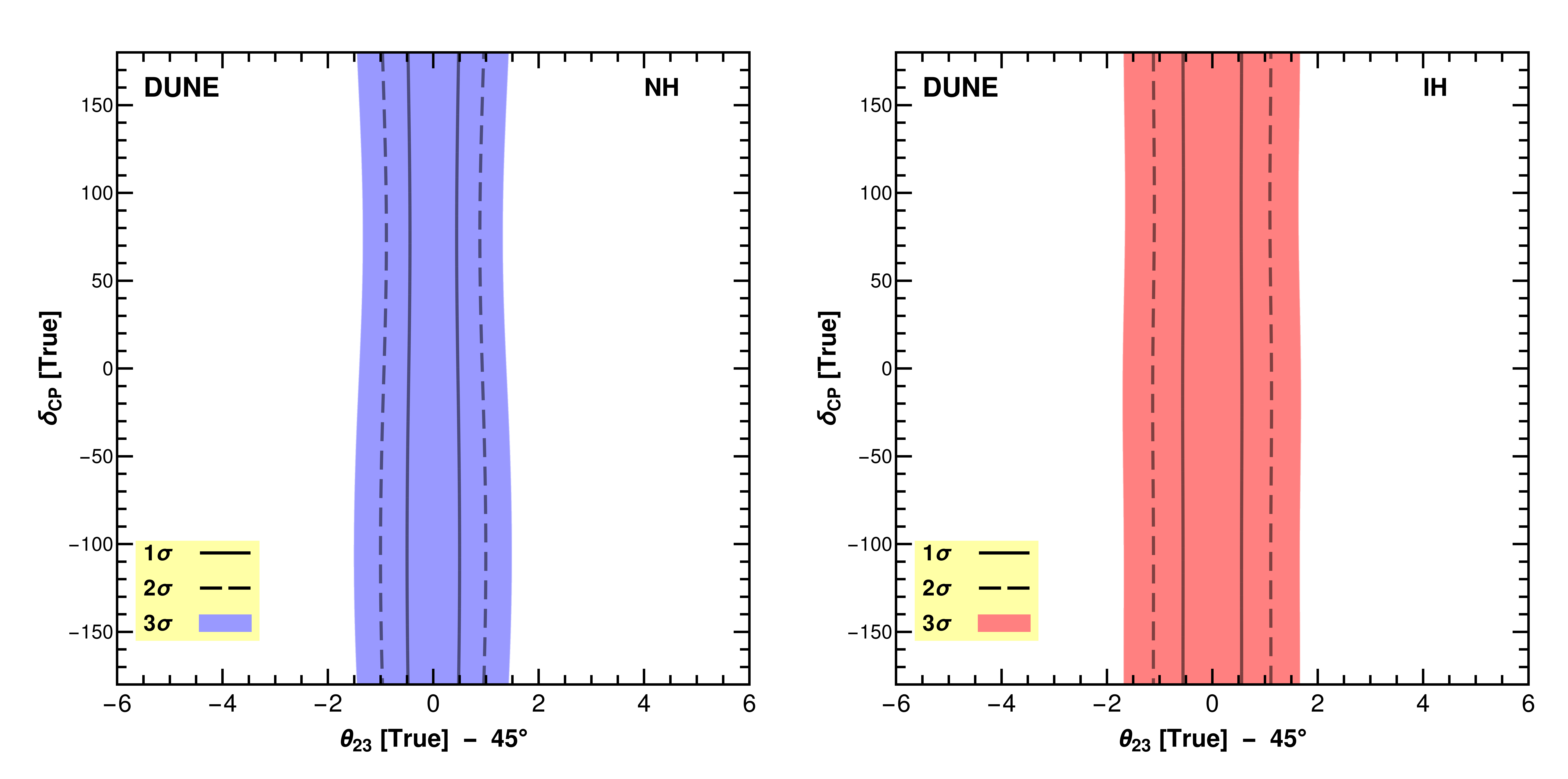}
\caption{\label{fig:1} Octant determination in DUNE under the standard three-neutrino paradigm. The white regions show the values of $\theta_{23}$ and $\delta_{CP}$ at which the octant of $\theta_{23}$ could be determined at a 3$\,\sigma$ CL or better. In the colored regions, conversely, the significance falls under 3$\,\sigma$. The 1$\,\sigma$ and 2$\,\sigma$ CL contours are shown with dashed and solid lines, and the sensitivities are presented for both NH (blue, left panel) and IH (red, right panel) mass orderings.}
\end{figure*}

In the standard model there exist three active neutrinos ($\nu_e$, $\nu_\mu$ and $\nu_\tau$) and no sterile neutrinos. Using the best-fit values and errors presented in Table \ref{bounds:1} we plotted the 1$\,\sigma$, 2$\,\sigma$ and 3$\,\sigma$ confidence level contours for various possible true values of $\theta_{23}$ and $\delta_{CP}$ in both NH and IH. The results are presented in Fig.\,\ref{fig:1}. We obtained this figure by keeping $\theta_{23}$ and $\delta_{CP}$ fixed to their assigned values, whilst the other four oscillation parameters ($\theta_{12}$, $\theta_{13}$, $\Delta m_{21}^2$, $\Delta m_{31}^2$) and the matter density were included in the $\chi^2$ minimization.

Fig.\,\ref{fig:1} is to be read as follows. The white regions correspond to the $\theta_{23}$ and $\delta_{CP}$ values where the octant of $\theta_{23}$ can be determined by DUNE at a 3$\,\sigma$ confidence level or better. In the colored region the sensitivity falls below 3$\,\sigma$, 2$\,\sigma$ or 1$\,\sigma$, where the latter two are indicated by the dashed and solid lines, respectively. In other words, if the values of $\theta_{23}$ and $\delta_{CP}$ fall in the region outside the band bordered by the dashed (solid) lines the octant of $\theta_{23}$ can be determined at a 2$\,\sigma$ (3$\,\sigma$) confidence level or better.

\subsection{\label{The nonunitary mixing case}The nonunitary mixing case}

In the case of nonunitary mixing, the sensitivity to the $\theta_{23}$ octant is evaluated by using Eq.\,(\ref{NP1}) or Eq.\,(\ref{NP2}) to calculate the mixing matrix and Eq.\,(\ref{HN}) to construct the corresponding Hamiltonian.

We take parameters $\alpha_{ij}$, $i,j=1,2,3$, and $\alpha_{\ell\ell'}$, $\ell,\ell'=e,\mu,\tau$, into account as Gaussian priors, where central values are set to match the standard three-neutrino case and the standard deviations are taken from the 1$\,\sigma$ bounds corresponding to the 90\% CL ones presented in Table \ref{bounds:2}.

\begin{figure*}
\includegraphics[width=\linewidth]{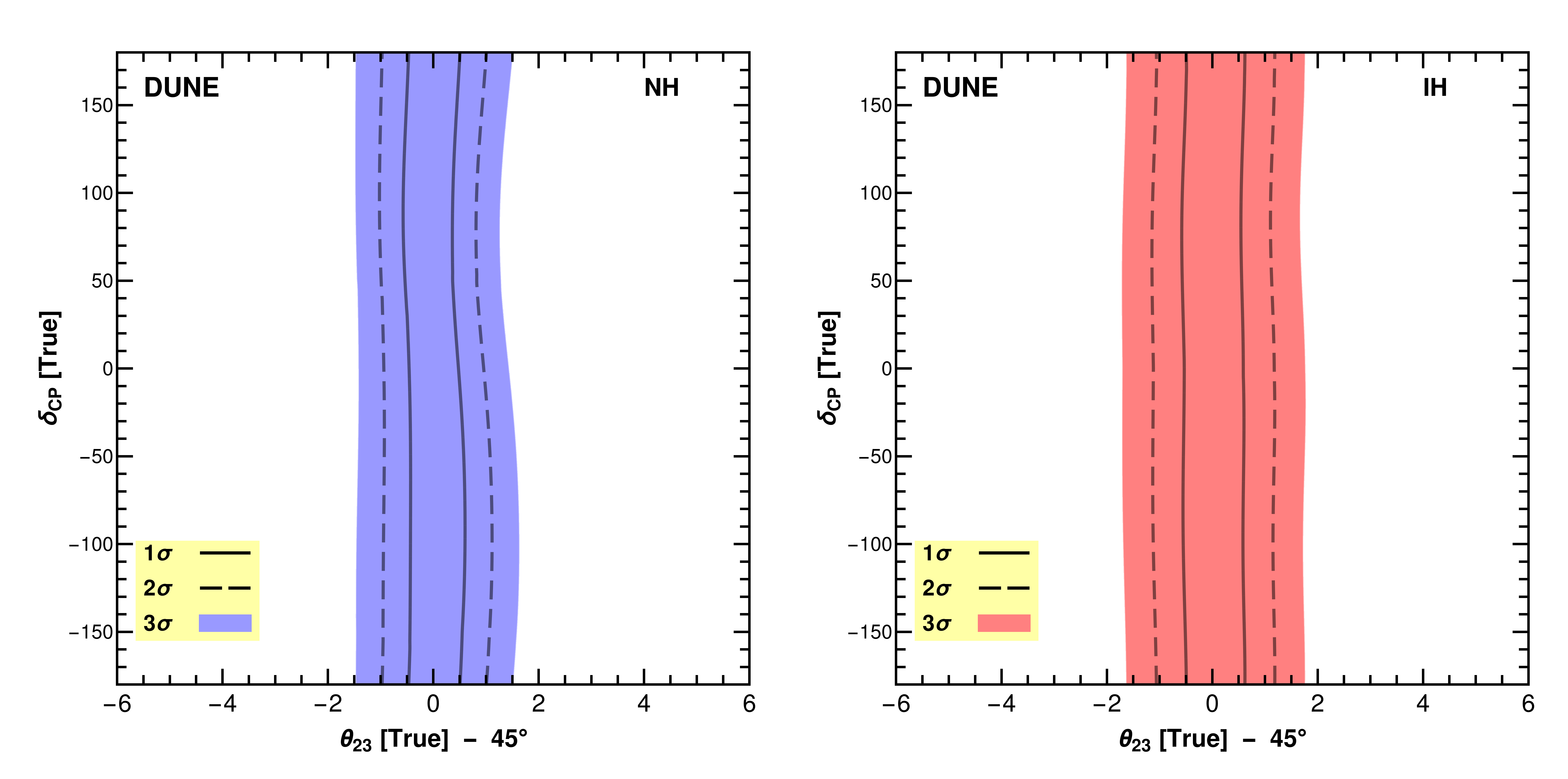}
\caption{\label{fig:2} Octant determination in DUNE under nonunitary mixing. The white regions show the values of $\theta_{23}$ and $\delta_{CP}$ at which the octant of $\theta_{23}$ could be determined at a 3$\,\sigma$ CL or better. In the colored regions, conversely, the significance falls under 3$\,\sigma$. The 1$\,\sigma$ and 2$\,\sigma$ CL contours are shown with dashed and solid lines, and the sensitivities are presented for both NH (blue, left panel) and IH (red, right panel) mass orderings.}
\end{figure*} 
In Fig.\,\ref{fig:2}, we present the 1$\,\sigma$, 2$\,\sigma$ and 3$\,\sigma$ contours for the octant determination under nonunitary mixing in the $\alpha_{\ell\ell'}$ basis. The sensitivity plots are shown both in the NH and IH cases. We also studied the sensitivities in the $\alpha_{\ell\ell'}$ basis, and found the results to be nearly identical to the ones obtained in the $\alpha_{ij}$ basis.

\subsection{\label{The light sterile neutrino case}The light sterile neutrino case}

\begin{figure*}
\includegraphics[width=\linewidth]{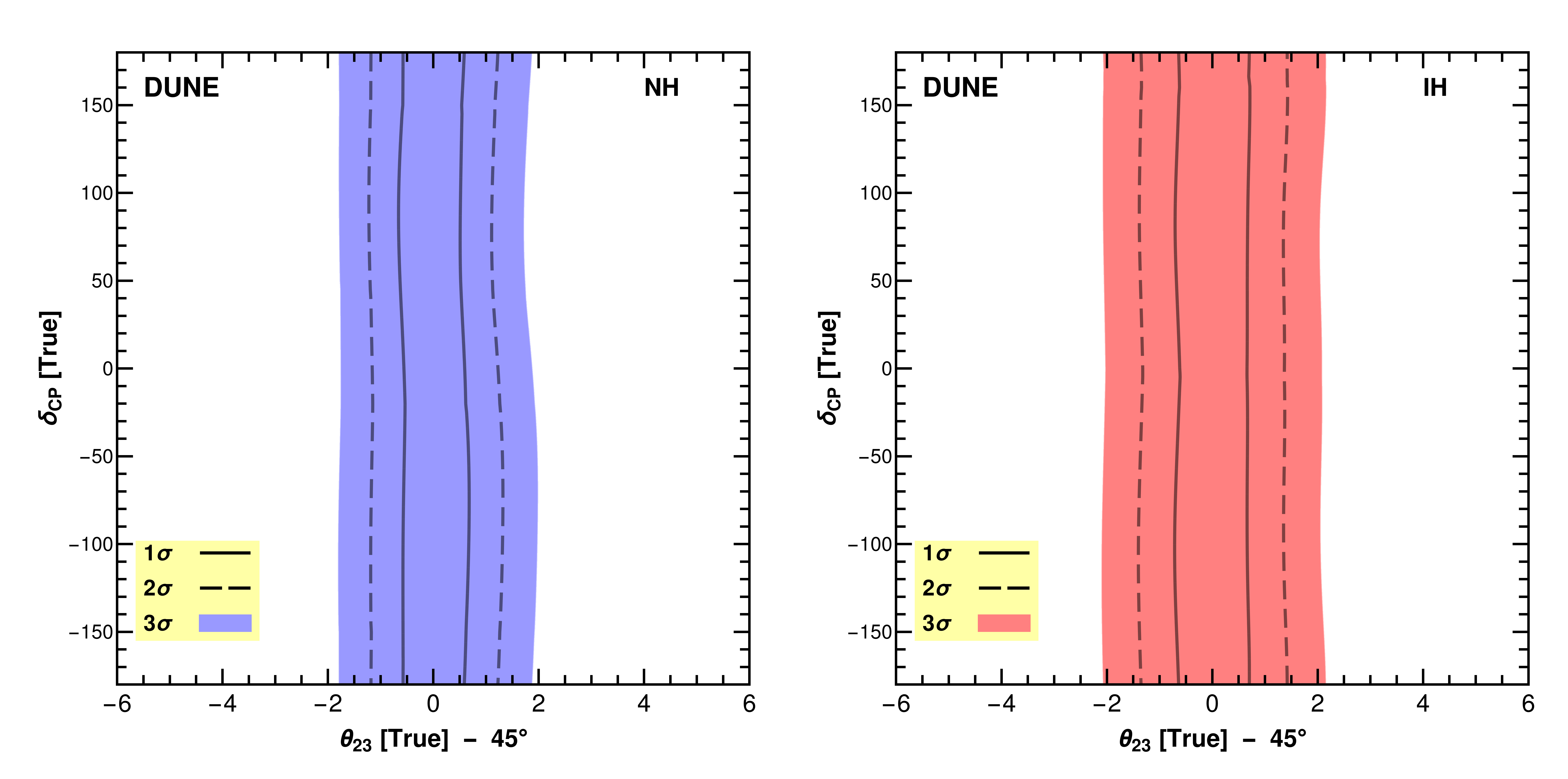}
\caption{\label{fig:3} Octant determination in DUNE in the presence of light sterile mixing with $\Delta m_{41}\geq 100$ eV$^2$. The white regions show the values of $\theta_{23}$ and $\delta_{CP}$ at which the octant of $\theta_{23}$ could be determined at a 3$\,\sigma$ CL or better. In the colored regions, conversely, the significance falls under 3$\,\sigma$. The 1$\,\sigma$ and 2$\,\sigma$ CL contours are shown with dashed and solid lines, and the sensitivities are presented for both NH (blue, left panel) and IH (red, right panel) mass orderings.}
\end{figure*}

Using the bounds derived for the mixing with a light sterile neutrino of $\Delta m_{41}^2\geq 100$ eV$^2$ mass range, shown in the centre column of Table \ref{bounds:3}, we obtained the sensitivity contours shown in Fig.\,\ref{fig:3}. We also studied the case of $0<\Delta m_{41}^2<1$ eV$^2$ by using the bounds presented in the right column of Table \ref{bounds:3}, but we did not find any significant difference to the $\Delta m_{41}^2\geq 100$ eV$^2$ case.

\subsection{\label{The unconstrained new physics case}The unconstrained new physics case}

Since it is unknown what may lie beyond the standard model, it is also necessary to discuss the scenario where none of the bounds that have been derived for nonunitary mixing or light sterile neutrinos may apply. An example of this situation could be a scenario where the sterile neutrinos are accompanied by nonstandard interactions (see e.g.\,\cite{Escrihuela:2016ube}) mediated by new Higgs particles, as is the case in left-right symmetric models. Due to its unknown nature, we consider an arbitrary form of new physics by calculating the $\chi^2$ values with no constraints on the $\alpha$ parameters.

\begin{figure*}
\includegraphics[width=\linewidth]{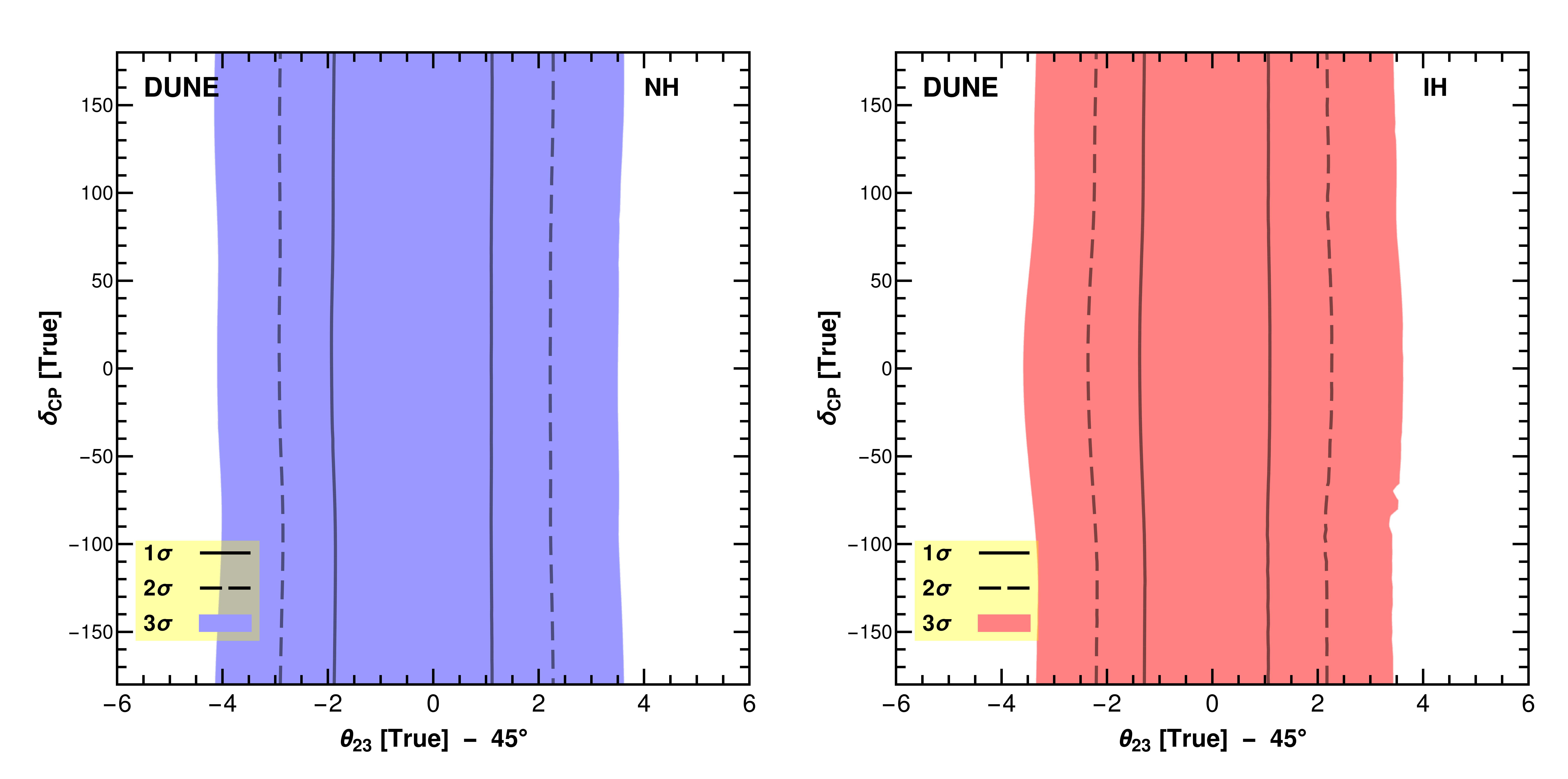}
\caption{\label{fig:4} Octant determination in DUNE with unconstrained $\alpha_{\ell\ell'}$ parameters. The white regions show the values of $\theta_{23}$ and $\delta_{CP}$ at which the octant of $\theta_{23}$ could be determined at a 3$\,\sigma$ CL or better. In the colored regions, conversely, the significance falls under 3$\,\sigma$. The 1$\,\sigma$ and 2$\,\sigma$ CL contours are shown with dashed and solid lines, and the sensitivities are presented for both NH (blue, left panel) and IH (red, right panel) mass orderings.}
\end{figure*}

We calculated the sensitivity of DUNE to the $\theta_{23}$ octant in DUNE after removing all priors that concern $\alpha_{ij}$ where, $i,j=1,2,3$. The results are presented in Fig.\,\ref{fig:4}. One notices that maximizing the effects of nonunitary and light sterile neutrinos roughly worsens the sensitivity of DUNE to the octant in terms of the angle.

\begin{figure*}
\includegraphics[width=\linewidth]{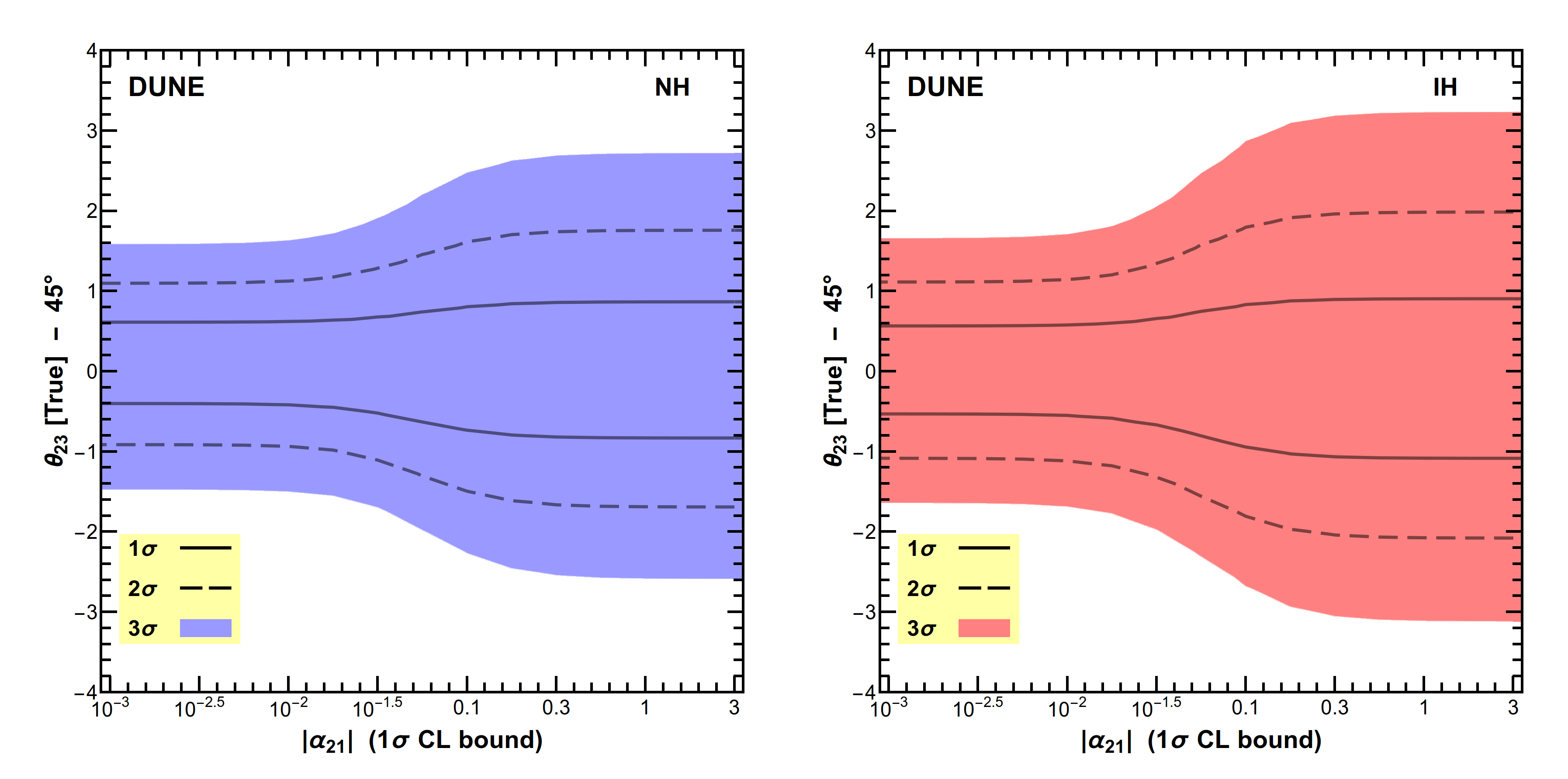}
\caption{\label{fig:5} Octant determination in DUNE as function of the 1$\,\sigma$ upper bound on $|\alpha_{21}|$. The phase of $\alpha_{21}$ is allowed to vary freely in the range $[0,2\pi]$ and the other alpha parameters are set to correspond to the standard three-neutrino paradigm.}
\end{figure*}

In order to get an understanding on how the magnitude of the $\alpha$ parameters affects the worsening of the sensitivity to the $\theta_{23}$ octant in the event where only $\alpha_{21}$ is taken into account, whereas the other alpha parameters are excluded from the $\chi^2$ calculation. In Fig.\,\ref{fig:5} we show the octant sensitivity as a function of the 1$\,\sigma$ upper bound of $|\alpha_{21}|$, and allow the phase of $\alpha_{21}$ vary freely in the range $[0,2\pi]$. Clearly, the contours in Fig.\,\ref{fig:5} show that the constraint on $\alpha_{21}$ affects the octant determination when $\alpha_{21}\gtrsim 10^{-2}$.

\section{\label{Conclusions}Conclusions}

We have presented the sensitivity to the determination of the $\theta_{23}$ octant ($\theta_{23}\leq\pi/4$ or $\theta_{23}\geq\pi/4$) in DUNE in four different scenarios. On the one hand, we have updated the 1$\,\sigma$, 2$\,\sigma$ and 3$\,\sigma$ confidence level contours for the standard model, where oscillations are constituted between three active neutrinos. On the other hand, we have also given these contours for three different scenarios where the octant sensitivity is interfered by sterile neutrinos and other potential sources for physics beyond the standard model. We analyzed these scenarios by parametrizing the new physics with the methods that were originally introduced in Refs.\,\cite{Escrihuela:2015wra} and \cite{Blennow:2016jkn} to describe nonunitarity of the light neutrino mixing matrix.

We found that the nonunitarity of the mixing matrix caused the sensitivity $\theta_{23}$ octant to decrease from the standard model case. Nevertheless, due to the strictness of the existing bounds for the nonunitarity parameters $\alpha_{ij}$, $i,j=1,2,3$ derived in Ref.\,\cite{Escrihuela:2016ube} and for $\alpha_{\ell\ell'}$, $\ell,\ell'=e,\mu,\tau$ derived in Ref.\,\cite{Blennow:2016jkn} the observed drop in the octant sensitivity was found to be very small. The worsening of the octant sensitivity due to sterile neutrino was found larger than this. The sensitivity was calculated in this case using the bounds on $\alpha_{\ell\ell'}$ given in Ref.\,\cite{Blennow:2016jkn}. The worsening of the sensitivity was found to be less than 1$^\circ$ in each octant.

We found the decrease in sensitivity due to the light sterile neutrino to be substantially less significant than that reported in Ref.\,\cite{Agarwalla:2016xlg} where the impact of a sterile neutrino with mixing angles $\theta_{14}=\theta_{24}=9^\circ$ and $\theta_{34}=0^\circ$ was considered in the determination of the $\theta_{23}$ octant in DUNE. Evidence of this sensitivity decrease can be seen from the comparison between our Fig.\,\ref{fig:1} with Fig.\,3 of Ref.\,\cite{Agarwalla:2016xlg}. When converted to the nonunitarity formalism (see the Appendix of Ref.\,\cite{Escrihuela:2015wra}), this kind of sterile neutrino would imply nonunitarity whose parameter values lie close to the existing bounds we presented for $0<\Delta m^2_{41}<1$ eV$^2$ in Table \ref{bounds:3}. On the other hand, our investigation takes into account all possibilities for light sterile neutrinos, whereas the authors of Ref.\,\cite{Agarwalla:2016xlg} consider a specific model. Thus our results are in this respect more general, therefore statistically favored by comparison and hence the difference between the two sets. If the model of Ref.\,\cite{Agarwalla:2016xlg} is realized in nature, then the ability of DUNE to tell the $\theta_{23}$ octancy is deteriorated.

We also tested how the octant sensitivity changed when the new physics parameters $\alpha_{ij}$ were left unconstrained. This type of simulation corresponds to a new physics scenario, where sterile neutrinos are associated with other new physics effects, not taken into account in Refs.\,\cite{Blennow:2016jkn} and \cite{Escrihuela:2016ube} when deriving the bounds for the nonunitary and light sterile mixing effects. An example of this could be nonstandard interactions involved in the neutrino propagation. Our simulations showed that in the worst case the octant could be determined at 3$\,\sigma$ CL or better for $\theta_{23}\lesssim 41.0^\circ$ and $\theta_{23}\gtrsim 48.5^\circ$ for the normal hierarchy to be compared with the bounds $\theta_{23}\lesssim 43.5^\circ$ and $\theta_{23}\gtrsim 46.5^\circ$ of the standard case.

In conclusion, we found that nonunitarity of the neutrino mixing matrix or the possible existence of light sterile neutrinos affect only mildly the sensitivity of DUNE to determine the octant of $\theta_{23}$. This is in contrast with the determination of the $CP$ violation, where the presence of sterile neutrinos could jeopardize the sensitivity \cite{Ge:2016xya, Escrihuela:2016ube, Blennow:2016jkn}.

After submitting our paper, we became aware of Ref.\,\cite{Tang:2017khg}, which covers partly the same topics we consider in this paper.

\begin{acknowledgments}
S.\,V.\,would like to express his gratitude for the Centro de F\'isica Te\'orica de Part\'iculas for the hospitality during his visit at the University of Lisbon, and the University of Jyv\"askyl\"a for a mobility grant that made this visit possible. C.\,R.\,D.\,thanks, Prof.\,Dmitri I.\,Kazakov, Director, BLTP, JINR for his kind support.
\end{acknowledgments}


\begin{thebibliography}{99}
\bibitem{Fogli:1996pv} G. L. Fogli and E. Lisi, Phys. Rev. D {\bf 54}, 3667 (1996), arXiv:hep-ph/9604415 [hep-ph].
\bibitem{Forero:2014bxa} D. V. Forero, M. Tortola, and J. W. F. Valle, Phys. Rev. D {\bf 90}, 093006 (2014), arXiv:1405.7540 [hep-ph].
\bibitem{Capozzi:2016rtj} F. Capozzi, E. Lisi, A. Marrone, D. Montanino, and A. Palazzo, Nucl. Phys. {\bf B908}, 218 (2016), arXiv:1601.07777 [hep-ph].
\bibitem{Gonzalez-Garcia:2015qrr}  M. C. Gonzalez-Garcia, M. Maltoni, and T. Schwetz, Nucl. Phys. {\bf B908}, 199 (2016), arXiv:1512.06856 [hep-ph].
\bibitem{Fogli:2012ua} G. L. Fogli, E. Lisi, A. Marrone, D. Montanino, A. Palazzo, and A. M. Rotunno, Phys. Rev. D {\bf 86}, 013012 (2012), arXiv:1205.5254 [hep-ph].
\bibitem{Adamson:2017gxd} P. Adamson {\it et al.} (NOvA), Phys. Rev. Lett. {\bf 118}, 231801 (2017), arXiv:1703.03328 [hep-ex].
\bibitem{Wendell:2010md} R. Wendell {\it et al.} (Super-Kamiokande), Phys. Rev. D {\bf 81}, 092004 (2010), arXiv:1002.3471 [hep-ex].
\bibitem{Adamson:2014vgd} P. Adamson {\it et al.} (MINOS), Phys. Rev. Lett. {\bf 112}, 191801 (2014), arXiv:1403.0867 [hep-ex].
\bibitem{Abe:2015awa} K. Abe {\it et al.} (T2K), Phys. Rev. D {\bf 91}, 072010 (2015), arXiv:1502.01550 [hep-ex].
\bibitem{Adamson:2016xxw} P. Adamson {\it et al.} (NOvA), Phys. Rev. D {\bf 93}, 051104 (2016), arXiv:1601.05037 [hep-ex].
\bibitem{Agarwalla:2013ju} S. K. Agarwalla, S. Prakash, and S. U. Sankar, J. High Energy Phys. {\bf 07}, 131 (2013), arXiv:1301.2574 [hep-ph].
\bibitem{Ghosh:2014rna} M. Ghosh, S. Goswami, and S. K. Raut, Eur. Phys. J. C {\bf 76}, 114 (2016), arXiv:1412.1744 [hep-ph].
\bibitem{Nath:2015kjg}  N. Nath, M. Ghosh, and S. Goswami, Nucl. Phys. {\bf B913}, 381 (2016), arXiv:1511.07496 [hep-ph].
\bibitem{Fukasawa:2016yue} S. Fukasawa, M. Ghosh, and O. Yasuda, Nucl. Phys. {\bf B918}, 337 (2017), arXiv:1607.03758 [hep-ph].
\bibitem{Ballett:2016daj} P. Ballett, S. F. King, S. Pascoli, N. W. Prouse, and T. C. Wang, Phys. Rev. D {\bf 96}, 033003 (2017), arXiv:1612.07275 [hep-ph].
\bibitem{Chatterjee:2017irl} S. Sachi Chatterjee, P. Pasquini, and J. W. F. Valle, Phys. Rev. D {\bf 96}, 011303 (2017), arXiv:1703.03435 [hep-ph].
\bibitem{Agarwalla:2017nld} S. K. Agarwalla, M. Ghosh, and S. K. Raut, J. High Energy Phys. {\bf 05}, 115 (2017), arXiv:1704.06116 [hep-ph].
\bibitem{Altarelli:2010gt} G. Altarelli and F. Feruglio, Rev. Mod. Phys. {\bf 82}, 2701 (2010), arXiv:1002.0211 [hep-ph].
\bibitem{CarcamoHernandez:2017owh}  A. E. Carcamo Hernandez, S. Kovalenko, J. W. F. Valle, and C. A. Vaquera-Araujo, J. High Energy Phys. {\bf 07}, 118 (2017), arXiv:1705.06320 [hep-ph].
\bibitem{Hanlon:2013ska} A. D. Hanlon, S.-F. Ge, and W. W. Repko, Phys. Lett. B {\bf 729}, 185 (2014), arXiv:1308.6522 [hep-ph].
\bibitem{Agarwalla:2016xlg} S. K. Agarwalla, S. S. Chatterjee, and A. Palazzo, Phys. Rev. Lett. {\bf 118}, 031804 (2017), arXiv:1605.04299 [hep-ph].
\bibitem{Agarwalla:2016fkh} S. K. Agarwalla, S. S. Chatterjee, and A. Palazzo, Phys. Lett. B {\bf 762}, 64 (2016), arXiv:1607.01745 [hep-ph].
\bibitem{Dutta:2016eks}  D. Dutta, P. Ghoshal, and S. K. Sehrawat, Phys. Rev. D {\bf 95}, 095007 (2017), arXiv:1610.07203 [hep-ph].
\bibitem{Choubey:2017cba} S. Choubey, D. Dutta, and D. Pramanik, Phys. Rev. D {\bf 96}, 056026 (2017), arXiv:1704.07269 [hep-ph].
\bibitem{Das:2016bwe} C. R. Das, J. Maalampi, J. Pulido, and S. Vihonen, J. Phys.: Conf. Ser. {\bf 888}, 012219 (2017), arXiv:1606.02504 [hep-ph].
\bibitem{Das:2014fja} C. R. Das, J. Maalampi, J. Pulido, and S. Vihonen, J. High Energy Phys. {\bf 02}, 048 (2015), arXiv:1411.2829 [hep-ph].
\bibitem{Hettmansperger:2011bt} H. Hettmansperger, M. Lindner, and W. Rodejohann, J. High Energy Phys. {\bf 04}, 123 (2011), arXiv:1102.3432 [hep-ph].
\bibitem{Escrihuela:2015wra} F. J. Escrihuela, D. V. Forero, O. G. Miranda, M. Tortola, and J. W. F. Valle, Phys. Rev. D {\bf 92}, 053009 (2015), [Erratum: Phys. Rev. D {\bf 93}, 119905 (2016)], arXiv:1503.08879 [hep-ph].
\bibitem{Blennow:2016jkn} M. Blennow, P. Coloma, E. Fernandez-Martinez, J. Hernandez-Garcia, and J. Lopez-Pavon, J. High Energy Phys. {\bf 04}, 153 (2017), arXiv:1609.08637 [hep-ph]. 
\bibitem{Huber:2004ka} P. Huber, M. Lindner, and W. Winter, Comput. Phys. Commun. {\bf 167}, 195 (2005), arXiv:hep-ph/0407333 [hep-ph].
\bibitem{Huber:2007ji} P. Huber, J. Kopp, M. Lindner, M. Rolinec, and W. Winter, Comput. Phys. Commun. {\bf 177}, 432 (2007), arXiv:hep-ph/0701187 [hep-ph].
\bibitem{Akhmedov:2004ny} E. K. Akhmedov, R. Johansson, M. Lindner, T. Ohlsson, and T. Schwetz, J. High Energy Phys. {\bf 04}, 078 (2004), arXiv:hep-ph/0402175 [hep-ph].
\bibitem{Nunokawa:2007qh} H. Nunokawa, S. J. Parke, and J. W. F. Valle, Prog. Part. Nucl. Phys. {\bf 60}, 338 (2008), arXiv:0710.0554 [hep-ph].
\bibitem{Ge:2016xya} S.-F. Ge, P. Pasquini, M. Tortola, and J. W. F. Valle, Phys. Rev. D {\bf 95}, 033005 (2017), arXiv:1605.01670 [hep-ph].
\bibitem{Escrihuela:2016ube} F. J. Escrihuela, D. V. Forero, O. G. Miranda, M. Tortola, and J. W. F. Valle, New J. Phys. {\bf 19}, 093005 (2017), arXiv:1612.07377 [hep-ph].
\bibitem{Acciarri:2015uup} R. Acciarri {\it et al.} (DUNE), (2015), arXiv:1512.06148 [physics.ins-det].
\bibitem{Alion:2016uaj} T. Alion {\it et al.} (DUNE), (2016), arXiv:1606.09550 [physics.ins-det]. 
\bibitem{Esteban:2016qun} I. Esteban, M. C. Gonzalez-Garcia, M. Maltoni, I. Martinez-Soler, and T. Schwetz, J. High Energy Phys. {\bf 01}, 087 (2017), arXiv:1611.01514 [hep-ph].
\bibitem{notefoot}Let us note that these values are not on any statistical level sensitive to the non-unitarity effects we consider \cite{Blennow:2016jkn}.
\bibitem{Tang:2017khg} J. Tang, Y. Zhang, and Y.-F. Li, Phys. Lett. B {\bf 774}, 217 (2017), arXiv:1708.04909 [hep-ph].
\end{thebibliography}
\end{document}